\documentclass[sn-mathphys,Numbered]{sn-jnl}

\usepackage{graphicx}%
\usepackage{multirow}%
\usepackage{amsmath,amssymb,amsfonts}%
\usepackage{amsthm}%
\usepackage{mathrsfs}%
\usepackage[title]{appendix}%
\usepackage{xcolor}%
\usepackage{textcomp}%
\usepackage{manyfoot}%
\usepackage{booktabs}%
\usepackage{algorithm}%
\usepackage{algorithmicx}%
\usepackage{algpseudocode}%
\usepackage{listings}%
\usepackage{adjustbox}
\usepackage{caption}
\usepackage{subcaption}
\usepackage{tabularx}

\theoremstyle{thmstyleone}%
\theoremstyle{thmstyletwo}%

\theoremstyle{thmstylethree}%

\begin{document}

\title{A Network-Based Measure of Cosponsorship Influence on Bill Passage in the United States House of Representatives}


\author{\fnm{Sarah} \sur{Sotoudeh$^1$}}

\author{\fnm{Mason A.} \sur{Porter$^{1,2,3}$}}

\author{\fnm{Sanjukta} \sur{Krishnagopal$^{4}$}}

\affil{\orgdiv{$^1$Department of Mathematics}, \orgname{University of California, Los Angeles}, \orgaddress{ \city{Los Angeles}, \state{California},  \postcode{90095}, \country{USA}}}

\affil{\orgdiv{$^2$Department of Sociology}, \orgname{University of California, Los Angeles}, \orgaddress{ \city{Los Angeles}, \state{California},  \postcode{90095}, \country{USA}}}

\affil{\orgdiv{$^3$}\orgname{Santa Fe Institute}, \orgaddress{ \city{Santa Fe}, \state{New Mexico},  \postcode{87501}, \country{USA}}}

\affil{\orgdiv{$^4$Department of Computer Science}, \orgname{University of California, Santa Barbara}, \orgaddress{ \city{Santa Barbara}, \state{California},  \postcode{93016}, \country{USA}}}


\abstract{Each year, the United States Congress considers thousands of legislative proposals to select bills to present to the US President to sign into law. Naturally, the decision processes of members of Congress are subject to peer influence. In this paper, we examine the effect on bill passage of accrued influence between US Congress members in the US House of Representatives. We explore how the influence of a bill's cosponsors affects the bill's outcome (specifically, whether or not it passes in the House). We define a notion of influence by analyzing the structure of a network that we construct using cosponsorship dynamics. We award `influence' between a pair of Congress members when they cosponsor a bill that achieves some amount of legislative success. We find that properties of the bill cosponsorship network can be a useful signal to examine influence in Congress; they help explain why some bills pass and others fail. We compare our measure of influence to off-the-shelf centrality measures and conclude that our influence measure is more indicative of bill passage.}


\keywords{Political Networks, United States Congress, United States House of Representatives, Cosponsorship Networks, legislative influence, centrality}


\maketitle
\newpage



\section{Introduction}\label{sec1}

Since 1789, members of the United States Congress have been entrusted to help determine the laws that shape the US's political landscape \cite{smith2013american}. Congress has two chambers: the House of Representatives, which has 435 legislators (which are called representatives); and the Senate, which has 100 legislators (which are called senators). Congress decides which pieces of legislation to draft, which bills to vote on, and ultimately which bills to pass to present to the US President. Congress considers thousands of legislative proposals each year, but only a few hundred of them ultimately pass and are enacted into law \cite{govtrackHistoricalStatistics}. Why do some bills pass but other bills fail?

Political scientists have long considered the role that social networks play in politics \cite{carpenter2004,victor2009,campbell2013social,klofstad2013,kirkland2014,reilly2017social}. Humans are social creatures, and the choices that they make are influenced by the people and environments around them. There is evidence that interpersonal relationships influence political behavior, including in voter decisions during presidential elections \cite{lazer2010,beck2002social}. Naturally, one also expects the behavior of politicians themselves to be subject to influence from their peers. In both the House of Representatives and the Senate, researchers have observed that party affiliation influences roll-call votes beyond legislators’ political preferences~\cite{minozzi2021congress, snyder2000estimating}. Accordingly, legislators are affected by the opinions of others in their social networks during their political decision-making processes. 

During the last two decades, many researchers have used network analysis to study social and political ties between legislators~\cite{lazer_2011,ward2011network,victor2016,battaglini2019social}. More generally, network analysis is an informative approach to examine the mutual dependencies and channels of influence between interacting individuals, groups, and institutions \cite{wasserman1994,newman2018}. It can thus yield important insights into political processes, which involve complex relationships between many entities. Researchers have analyzed many types of political and social networks. These include joint membership of legislative committees and subcommittees \cite{porter2005network,porter2007}, digital-trace data on social media \cite{peng2016follower}, coappearances in World Wide Web searches \cite{sanghoon2010}, and bill cosponsorship networks~\cite{fowler2006connecting,fowler2006b}.

An important application of networks in political science is the analysis of how groups of individuals debate, decide, and commit to decisions (such as reaching a verdict as a jury, allocating funds as a committee, or a passing a bill in Congress \cite{victor2016}). The analysis of bill cosponsorship networks can yield insight into the effects of relationships and networks on political processes~\cite{fowler2006connecting,fowler2006b}. A proposed bill in Congress has one sponsor (the main member of Congress that introduces the bill) and can have any number (even a very large number) of cosponsors. Cosponsorship data captures a notion of collaboration, and it can also indicate information about the time evolution of relationships and social-tie strengths between Congress members. Many researchers have analyzed cosponsorship networks to obtain insights into political dynamics~\cite{fowler2006connecting,fowler2006b,zhang2008,NEAL202297,Clark_Caro_2013,neal2020sign}. For example, Fowler~\cite{fowler2006connecting} formulated and analyzed a `connectedness' centrality measure to identify important members of Congress in bill cosponsorship networks. This connectedness measure considers a weighted count --- using both the numbers of cosponsors on bills and the numbers of bills that pairs of legislators both cosponsor --- of bill cosponsorship between legislators, takes the reciprocal of this weight to yield a cosponsorship distance between each two legislators, and uses Dijkstra's algorithm to find the shortest distance between each pair of legislators. Fowler then calculated a notion of connectedness for each legislator using that legislaor's mean distance to the other legislators. Fowler illustrated that such connectedness can help successfully infer the passage of amendments to existing legislation, thereby using properties of bill cosponsorship networks as a proxy for legislator influence.

In the present paper, we define a notion of `influence' between Congress members that depends on achieving some level of legislative success. To examine such legislator influence, we consider bill cosponsorship relationships between members of the US House of Representatives. The incorporation of legislative outcomes into our notion of influence is a key distinction between our approach and the connectedness measure in~\cite{fowler2006connecting}. We find that our notion of influence is a better predictor of bill passage than several conventional network centrality measures (specifically, eigenvector centrality, closeness centrality, and strength centrality).

Our paper proceeds as follows. In Section \ref{sec2}, we contextualize bill cosponsorship in the US Congress and detail its relevance as a potential indicator of legislator influence. In Section \ref{sec3}, we discuss the data (including our preprocessing steps) that we employ in our investigation. In Section \ref{sec4}, we calculate influence in the US House of Representatives using bill cosponsorship data from January 2009 through January 2019 (i.e., the 111th--115th Congresses). In Section \ref{sec5}, we compare our notion of influence to eigenvector centrality. We conclude in Section \ref{sec6} by discussing our results and outlining further research directions. In Appendix \ref{secA2}, we compare our notion of influence to closeness centrality and strength centrality. Our code is publicly available at \url{https://github.com/sarahsotoudeh/LegislativeInfluence}. 


\section{Cosponsorship History and Relevance of Cosponsorship to Legislator Influence}\label{sec2}

More than half of the bills that are introduced in the US Congress have at least one cosponsor \cite{wilson1997cosponsorship}. Bill cosponsorship, which began in the US in the 91st Congress (1969--1971) with a maximum of 25 cosponsors permitted per bill at the time \cite{congressionalbillsCongressionalBills}, offers a legislative tool for Congress members to present their political opinions without flooding Congress with sponsored bills. In the 96th Congress (1979--1981), the restriction on the number of cosponsors per bill was removed, allowing cosponsorship to become a more prominent mechanism for Congress members to attempt to garner and express support for bills \cite{congressionalbillsCongressionalBills}. 

Although bill cosponsorship is a common practice in the US Congress, scholars and pundits debate its significance in the legislative process~\cite{wilson1997cosponsorship}. Some researchers have argued that cosponsorship is a low-risk way for Congress members to signal to their constituents that they are following through on their campaign promises \cite{wilson1997cosponsorship}. If a bill that a Congress member cosponsors passes, they can use it as evidence to convey to their constituents that they are doing good work. If a bill that they cosponsor does not pass, which is common in Congress, there are no real consequences. Researchers have also found that the sheer number of cosponsors on a bill is not a strong indicator of its success \cite{wilson1997cosponsorship}, so perhaps bill cosponsorship does not play a significant role in the legislative process.  A more nuanced perspective is that cosponsorship ties can indicate the evolving political relationships between members of Congress, which ultimately may play a role in legislation, even if cosponsor count itself does not give a clear signal of a bill's success \cite{fowler2006connecting}.

Making decisions on which bills to cosponsor is a complex and nuanced process, and a Congress member may cosponsor a bill for a variety of reasons~\cite{koger2003position}. For example, they can support a bill because they believe that it is genuinely beneficial, because it appeases their constituents and may thus benefit their chances of reelection, or because it helps them curry favor with other legislators. In the 111th--115th Congresses (i.e., between January 2009 and January 2019), which is the data that we study (see Section \ref{sec3}), {the overwhelming majority of bills had at least one cosponsor. In the 111th--115th Congresses, at least 87.85\% of the bills in the House of Representatives had cosponsors in each of the 10 terms during these 5 Congresses. During the same period, at least 86.33\% of the bills in the Senate had cosponsors in each of the 10 terms.

Because support from other people is important in Congress, it seems likely that legislator decision-making is affected by a desire to make or maintain personal relationships, which may provide social capital for the future. Accordingly, a Congress member may be more likely to cosponsor a bill with an individual with whom they have a successful relationship through bill passage \cite{bratton2011networks}. Therefore, repeated bill cosponsorship between two or more Congress members may give evidence of legislative alliances, with patterns of supporting each other’s legislation \cite{BRANDENBERGER2018238}. Such \emph{quid pro quo} is sometimes called `logrolling' in legislative and other organizational contexts \cite{smith1996}. There is empirical evidence for such support patterns, with the structure of some social networks in Congress resembling the network structure in the mutual licking of cows \cite{faust2002comparing}. Legislator influence can also arise through homophily~\cite{NEAL202297}. For example, collaborations within a party occur not only from mere party loyalty (and the ensuing promotion of a party's agenda and `brand name') but also from ideological similarities \cite{jenkins2008}. Some legislators also cosponsor bills across party lines (especially for Senators, where it is helpful to advance career goals \cite{rippere2016}), even amidst growing partisanship in Congress~\cite{andris2015}. Drawing from these ideas, we argue that the passage of a large percentage of bills that a Congress member supports (as demonstrated through cosponsorship) may be an indication that that Congress member is influential.  


\section{Data and Preprocessing}\label{sec3}

We extract cosponsorship activity on proposed legislation in the US House of Representatives between January 2009 and January 2019 (i.e., the 111th--115th Congresses). Congressional elections occur every two years, so our data spans five Congresses. This data is publicly available through ProPublica, which retrieved the data from the US Congress \cite{ProPublica_2023}. To match Congress members who served multiple terms, we leverage other data sets to complete missing identification numbers in our main cosponsorship data set \cite{Congress_Legislators,Congressional_Committees,Messy_Congressional_Data}. In our analysis, we only consider bills; we exclude all other types of legislation (namely, simple resolutions, concurrent resolutions, and joint resolutions). Bills undergo a different legislative process than other types of legislation, so it is difficult to directly compare their success to the success of other types of legislation \cite{LOC2024,GovInfo_2023b}. Bills typically propose significant pieces of legislation; if successful, they carry the force of law, which is not the case of simple and concurrent resolutions. Joint resolutions are often used for continuing or emergency appropriations, so Congressional relationships are likely to have a different relevance for them than for bills. Bills constitute about 70\% of all proposed legislation in Congresses 111--115~\cite{ProPublica_2023}, so using only bills leaves us with ample data to analyze. 

In Figure \ref{billprocess}, we illustrate the life cycle of a proposed bill \cite{sullivan2007our}. A bill is drafted in either the Senate or the House of Representatives. If a bill passes in its originating chamber of Congress, it is sent to the other chamber for review. In our investigation, we only examine whether or not bills pass in their original chamber. We study bills in the US House of Representatives, and we say that a bill `passes' if it succeeds at the introduction, committee, and debate stages in Figure \ref{billprocess} (irrespective of whether or not it passes in the Senate and of whether or not it is enacted into law). We say that a bill `fails' if it fails at some stage in the House. We limit our analysis in this way for two primary reasons. First, Congress members can only cosponsor bills that are introduced in their own chamber \cite{sullivan2007our}. Therefore, it seems reasonable to study the effects of cosponsorship within a single Congressional chamber. Second, very few bills survive the legislative process to be enacted into law. Roughly 3\% of all bills {in Congresses 111--115 in} the ProPublica data (which covers both the House and the Senate) were ultimately enacted into law~\cite{ProPublica_2023}.

\begin{figure}[!htb]
\centering
\includegraphics[scale=0.38] {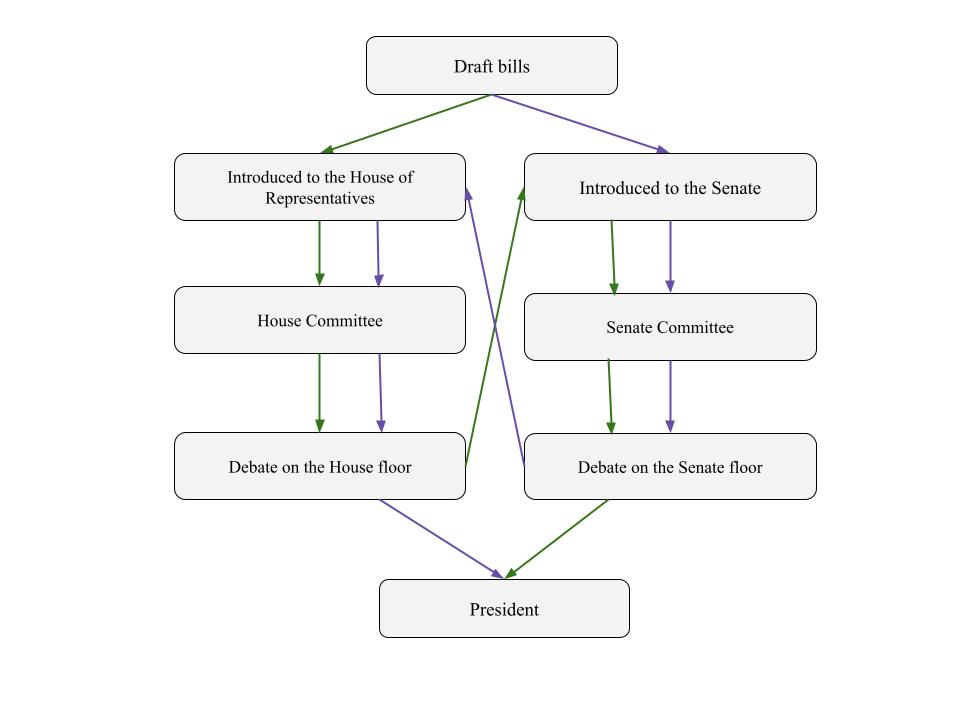}
\hspace*{100mm}
\vspace{-1.0 cm}
\caption{
{\footnotesize
\textbf{A simplistic overview of the legislative process in the United States Congress.} A bill starts in the drafting stage of a chamber and then trickles down in that chamber. If a bill passes in its originating chamber, it is introduced in the other chamber and goes through a similar review and voting process. Finally, if the bill passes both chambers, it is sent to the President, who either signs or vetoes the bill. The green and purple arrows indicate the paths of bills that originate in the House of Representatives and Senate, respectively. A bill can fail at any stage of this process.}
}
\label{billprocess}
\end{figure}

There are differences in cosponsorship behaviors in the Senate and the House of Representatives \cite{CosponsorshipConvo}. The House of Representatives (which has 435 representatives) is much larger than the Senate (which has 100 senators). Therefore, the House seemingly has more opportunities for cosponsorships and perhaps has a greater incentive to form political relationships. Representatives serve two-year terms, whereas senators serve six-year terms \cite{sullivan2007our}. Shorter terms entail more frequent reelection pressures and hence can hold individuals more accountable to their constituents. Unsurprisingly, cosponsorship activity increases dramatically in election years \cite{fowler2006connecting}. Additionally, because there are more legislators in the House than in the Senate, it may be more likely for a legislator in the former chamber to find a group of individuals with whom they align well politically.  

In Table \ref{below}, we summarize key statistics from the employed data set~\cite{ProPublica_2023} on bill passage in the House of Representatives.

\renewcommand{\arraystretch}{1.75}
\begin{table}[h]
\centering
\begin{tabularx}{\textwidth}{|l|*{5}{>{\centering\arraybackslash}X|}}
\hline
\multicolumn{1}{|c|}{} & \multicolumn{5}{c|}{{Congress}} \\
\multicolumn{1}{|c|}{} & \multicolumn{1}{c}{111} & \multicolumn{1}{c}{112} & \multicolumn{1}{c}{113} & \multicolumn{1}{c}{114} & \multicolumn{1}{c|}{115} \\
\hline
Percent of Bills that Pass in the House & 10.98\% & 7.3\% & 9.8\% & 11.84\% & 13.49\% \\
\hline
Percent of Bills that are Enacted into Law & 3.87\% & 2.96\% & 3.55\% & 3.31\% & 3.84\% \\
\hline
Mean Number of Cosponsors Per Bill & 21.50 & 21.44 & 21.96 & 20.93 & 19.01 \\
\hline
Maximum Number of Cosponsors Per Bill & 419 & 425 & 379 & 332 & 381 \\
\hline
Mean Number of Bills Per Cosponsor & 235.72 & 205.63 & 227.45 & 243.53 & 250.87 \\
\hline
Maximum Number of Bills Per Cosponsor & 797 & 632 & 666 & 695 & 791 \\
\hline
\end{tabularx}
\vspace{.3cm}
\caption{
{\footnotesize
{\bf Summary of data for bill passage in the US House of Representatives in Congresses 111--115.} In the first row, we show the percent of bills originating in the House of Representatives that pass in the House. In the second row, we show the percent of bills originating in the House of Representatives that were enacted into law. In the third row, we show the mean number of cosponsors of a bill. In the fourth row, we show the maximum number of cosponsors of a bill. In the fifth row, we show the mean number of bills that a representative cosponsors. In the sixth row, we show the maximum number of bills that a representative cosponsors.}}
\label{below}
\end{table}


\section{Defining Legislator Influence}\label{sec4}

We seek to define an influence score to capture the value of each representative's support on a bill based on (1) the historical success of their sponsored and cosponsored bills and (2) their bill sponsorship and cosponsorship relationships with other representatives.

To construct our influence score, we consider the influence that each representative has both on and from all other representatives at some time $t$. We examine such influence using a bill cosponsorship network in which each representative is a node and an undirected edge exists between representatives $i$ and $j$ if they sponsor or cosponsor a bill. We do not include any self-edges. We treat sponsorship and cosponsorship of bills on equal footing, rather than considering directed edges from bill cosponsors to sponsors. For simplicity, we use the term `cosponsor' to refer to any individual that either sponsors or cosponsors a bill. Our cosponsorship network is weighted, with time-dependent edge weights. The weight of an edge at time $t$ depends on the number of bills that two representatives both cosponsor at all times up to and including $t$, with decay as cosponsorship recedes into the past. In Section~\ref{sec4}, we give a detailed discussion of how we determine edge weights.

 To obtain a tensor representation of our bill cosponsorship network, we first construct
 an adjacency tensor with entries $A[t, i, j]_{\text{pass}}$ that encode
  the number of bills that include both representatives $i$ and $j$ as cosponsors,
  are introduced in month $t$, and eventually pass (including possibly in month $t$) in the House of Representatives.
  We also construct an adjacency tensor with entries $A[t, i, j]_{\text{tot}}$ that encodes the total number of bills, irrespective of their success, that include both representatives $i$ and $j$ as cosponsors and
  are introduced in month $t$. We suppose that representatives accumulate influence together when the bills that they mutually cosponsor pass. Accordingly, we assume that the influence between representatives $i$ and $j$ is symmetric. 
  Therefore, $A[t, i, j]_{\text{pass}}$ = $A[t, j, i]_{\text{pass}}$ and $A[t, i, j]_{\text{tot}} = A[t, j, i]_{\text{tot}}$. In future work, it is desirable to also consider asymmetric influence between sponsors and cosponsors.


\subsection{Accounting for the Accumulation and Decay of Influence with Time}
\label{subsec1}

We seek a cumulative-in-time notion of influence that incorporates decay with time. Intuitively, a legislator that has been in Congress for ten years has more `influence capital' (i.e., social capital for the purpose of influence) than a legislator on their first day in Congress \cite{goodwin1959seniority}. Additionally, a legislator that was very influential many years ago that has been inactive for a long time likely has experienced decay in their influence capital. To account for both of these ideas, we use an exponential decay function $e^{kt}$ so that accrued influence (from successful bills) decays as a function of the time $t$ since the introduction of a bill. The decay rate $k$ controls the half-life of a unit of influence between individuals that have successfully cosponsored a single passed bill. We examine influence half-lives of 6, 12, and 24 months by setting $k = \frac{\text{ln}(0.5)}{\text{half-life}} {\approx} \frac{-0.693}{\text{half-life}}$. 

We consider an exponential influence decay for both $A[t, i, j]_{\text{pass}}$ (the number of bills that are introduced in the House at time $t$, include both representatives $i$ and $j$ as cosponsors, and eventually pass in the House) and 
$A[t, i, j]_{\text{tot}}$ (the number of bills, irrespective of their success, that are introduced in the House at time $t$ and include both representatives $i$ and $j$ as cosponsors). We then aggregate results month by month to yield tensors $C_{\text{pass}}$ and $C_{\text{tot}}$ with entries
\begin{align}
    C[t, i, j]_{\text{pass}} &= \sum_{n = 1}^{t} {e^{k  (t - n)}}  A[n, i, j]_{\text{pass}}\,, \nonumber \\
    C[t, i, j]_{\text{tot}} &= \sum_{n = 1}^{t} {e^{k  (t - n)}} A[n, i, j]_{\text{tot}}\,{.}
    \label{basic_matrix}
\end{align}
The entry $C[t, i, j]_{\text{pass}}$ is the number of bills that are cosponsored by both representatives $i$ and $j$ and pass in the House between months $1$ and $t$ (inclusive of months $1$ and $t$), with a separate decay applied each month. 
For example, we obtain the influence between representatives $i$ and $j$ in month 3 (while accounting for the influence that they accumulate in months $1$ and $2$) by calculating
\begin{equation}
	C[3,i,j]_{\text{pass}} = e^{{2k}} A[1, i, j]_{\text{pass}} + e^{{1k}} A[2, i, j]_{\text{pass}} + e^{{0k}} A[3, i, j]_{\text{pass}} \,,
\end{equation}
which incorporates decay each month. Analogously, the entry $C[t, i, j]_{\text{tot}}$ is the total number of bills (irrespective of their success) that are cosponsored by both representatives $i$ and $j$ between months $1$ and $t$, with decay each month.


\subsection{Accounting for Political Parties}
\label{subsec2}

We account for legislators' political parties by calculating the influence of legislators with Republicans and Democrats separately. We control for party affiliation by separately summing across the Democrat and Republican columns in $C_{\text{pass}}$ and $C_{\text{tot}}$. We then divide the Democrat and Republican sums by the number of Democrats and number of Republicans, respectively, in the House of Representatives at time $t$ to examine the relative influence that representative $i$ has with Democrats and Republicans. This yields tensors with entries 
\begin{align}
    P[t, i]_{\text{{D}ems}} &= \frac{1}{N_{\text{Dems}}} \cdot \frac{\sum_{d} C[t, i, d]_{\text{pass}}}{\sum_{d} C[t, i, d]_{\text{tot}}}\,, \,\,\, d \in {\text{\{Democrats\}}} \,, \,\, {N}_{\text{Dems}} \, = |{\text{\{Democrats\}}}| \, , \nonumber \\
\vspace{6mm}
    P[t, i]_{\text{{R}eps}} &= \frac{1}{N_{\text{Reps}}} \cdot \frac{\sum_{d} C[t, i, r]_{\text{pass}}}{\sum_{d} C[t, i, r]_{\text{tot}}}\,, \,\,\, r \in {\text{\{Republicans\}}} \,,\,\, {N}_{\text{Reps}} \, = |{\text{\{Republicans\}}}| \,,
\end{align}
where $N_\text{Dems}$ and $N_\text{{R}eps}$, respectively, are the numbers of Democrats and Republicans in the House of Representatives at time $t$. The quantity $P[t, i]_{\text{{D}ems}}$ is the influence score of representative $i$ with Democrats 
in month $t$, and $P[t, i]_{\text{{R}eps}}$ is the influence score of representative $i$ with Republicans in month $t$. 

To compute a single scalar influence score for each representative, we sum $P[t, i]_{\text{{D}ems}}$ and $P[t, i]_{\text{{R}eps}}$ to obtain
\begin{align}
	I[t, i] = P[t, i]_{\text{{D}ems}} + P[t, i]_{\text{{R}eps}} \,{.}
\end{align}
The influence score $I[t, i]$ indicates the combined influence that representative $i$ has with Democrats and Republicans at time $t$.

\begin{figure}[H]
\centering
\includegraphics[scale=0.6] {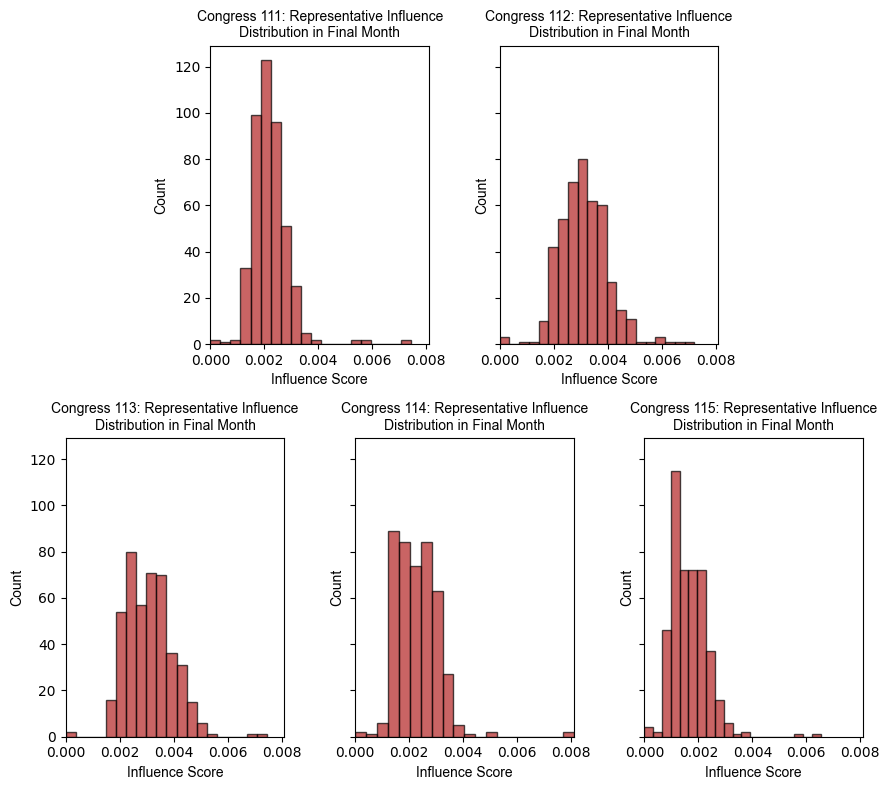}
\hspace*{100mm}
\caption{
{\footnotesize
{\textbf{Distribution of individual legislators in the US House of Representative in the final month of each Congress.}} 
} }
\label{inf_distribution}
\end{figure}

As we can see in Figure \ref{inf_distribution}, the distribution of the influence scores of individual legislators in the House of Representative ranges between $0$ and $0.008$. Perhaps unintutively, the mean of this distribution does not increase with time. In fact, the distribution mean seems to be smallest in Congresses 114 and 115. We hypothesis that this decrease in the distribution mean from the mean in Congresses 111--113 may be a signal of less collaboration between representatives in Congresses 114--115 than in the prior three Congresses. Congresses 114--115 coincide with the years 2015--2019, which saw a documented increase in political polarization in the US Congress~\cite{jacobson2018congress}.


\subsection{Computing Influence at the Bill Level}\label{subsec2}

To associate our measure of influence with bill-passage outcome, we examine influence at the bill level using time series of the mean and maximum cosponsor influence. The influence score of a bill aggregates the influence scores of its sponsors and cosponsors. We do this aggregation in two ways. In our first approach, we calculate a bill's influence score as the mean of the influence scores of its cosponsors in the month that the bill is introduced. In our second approach, we calculate a bill's influence score as the maximum of the influence scores of its cosponsors in the month that the bill is introduced. The two influence scores of bill $k$ are thus
\begin{align} \label{bill-influence}
	\text{Inf}_{\text{mean}, k} = \frac{\sum_{i \in \mathcal{C}_k} I[t_{k},i]}{| \mathcal{C}_k| } \,, \quad
		{{\text{Inf}_{\text{max}, k} = \max_{i \in  \mathcal{C}_k}{I[t_{k},i] }}} \,,
\end{align}
where $\mathcal{C}_k$ is the set of cosponsors of bill $k$, the number of cosponsors of bill $k$ is $|\mathcal{C}_\text{k}|$, and $t_{k}$ is the month that bill $k$ is introduced.

Our two bill-level influence scores capture two different ideas of legislator influence. The mean assigns equal weights to each cosponsor, while ensuring that a large number of cosponsors does not artificially inflate a bill's influence score. The maximum supposes that a bill's most influential cosponsor carries it and thereby garners a large amount of support for it.

Because we determine bill influence from legislator influence, which depends on the bill-passage outcome of many bills, whether or not a specific bill passes exerts only a small influence on its own influence score. We aggregate our results at a monthly level, whereas legislators vote on a bill on a specific day. In our aggregation, influence must either (1) include a bill in its own calculation or (2) not consider influence that arises from an earlier day of the same month. The first option can only change a numerator in Eq.~\eqref{bill-influence} by a maximum of 1 bill and the second option can yield a much more substantial change, so we use the first option. Additionally, these numerators are fairly large, so an increase of it by 1 bill has little effect on the calculation.


\section{Bill-Level Influence as an Indicator of Success in Bill Passage}\label{sec5}

We now compare our two bill-level influence scores and examine if either of them is indicative of whether a bill passes in the House of Representatives.


\subsection{Visualizing Bill-Level Influence}\label{subsec5.1}

We now illustrate how bill-level influence changes with time. These changes in bill-level influence reflect aggregate changes in cosponsor influence with time (see Eq.~\eqref{bill-influence}).

In the left panels of Figure \ref{basicinf}, we plot our bill-level influence scores as a function of time across nonoverlapping 4-month periods for bills that pass (orange crosses) and bills that fail (blue disks). We show bill-level influence scores from mean cosponsor influence in Figure \ref{basicinf}a, and we show the bill-level influence scores from maximum cosponsor influence in Figure \ref{basicinf}b. The corresponding histograms in the right panels show the distributions of the differences in mean bill-level influence between passed and failed bills (i.e., the `relative-difference distributions' between passed and failed bills) in the House. For both mean cosponsor influence and maximum cosponsoe influence, we calculate these relative differences by determining the difference between the mean bill-level influence scores of passed and failed bills and dividing that difference by the mean influence score of failed bills. This yields relative differences between passed and failed bills, and it thereby gives a normalized way to measure how much the bill-level influence of passed bills exceeds the bill-level influence of failed bills.

We use 4-month time periods as our (nonoverlapping) aggregation time windows for two key reasons. First, there are some months (especially near the end of each year) in which the House of Representatives passes very few bills. Aggregating data over a few months thus smooths out temporally localized fluctuations. Second, in the employed data, bills took a median time of 98 days (i.e., 3.2 months) and a mean time of 148 days (i.e., 4.9 months) between the time that they were introduced to the time that they passed in the House. Bills took a longer time to pass in the Senate, with a median time of 148 days (i.e., 4.9 months) and a mean time of 196 days (i.e., 6.5 months). Our choice of 4 months as an aggregation time is informed by the median and mean passage times in the House. Naturally, other choices are also viable, and future work can explore the effect of aggregating influence over different amounts of time. One can also consider overlapping time windows.

\begin{figure}[H]
    \centering
    
    \begin{subfigure}{\textwidth}
        \centering
        \adjustbox{max width=\textwidth, max height=0.5\textheight}{%
            \includegraphics{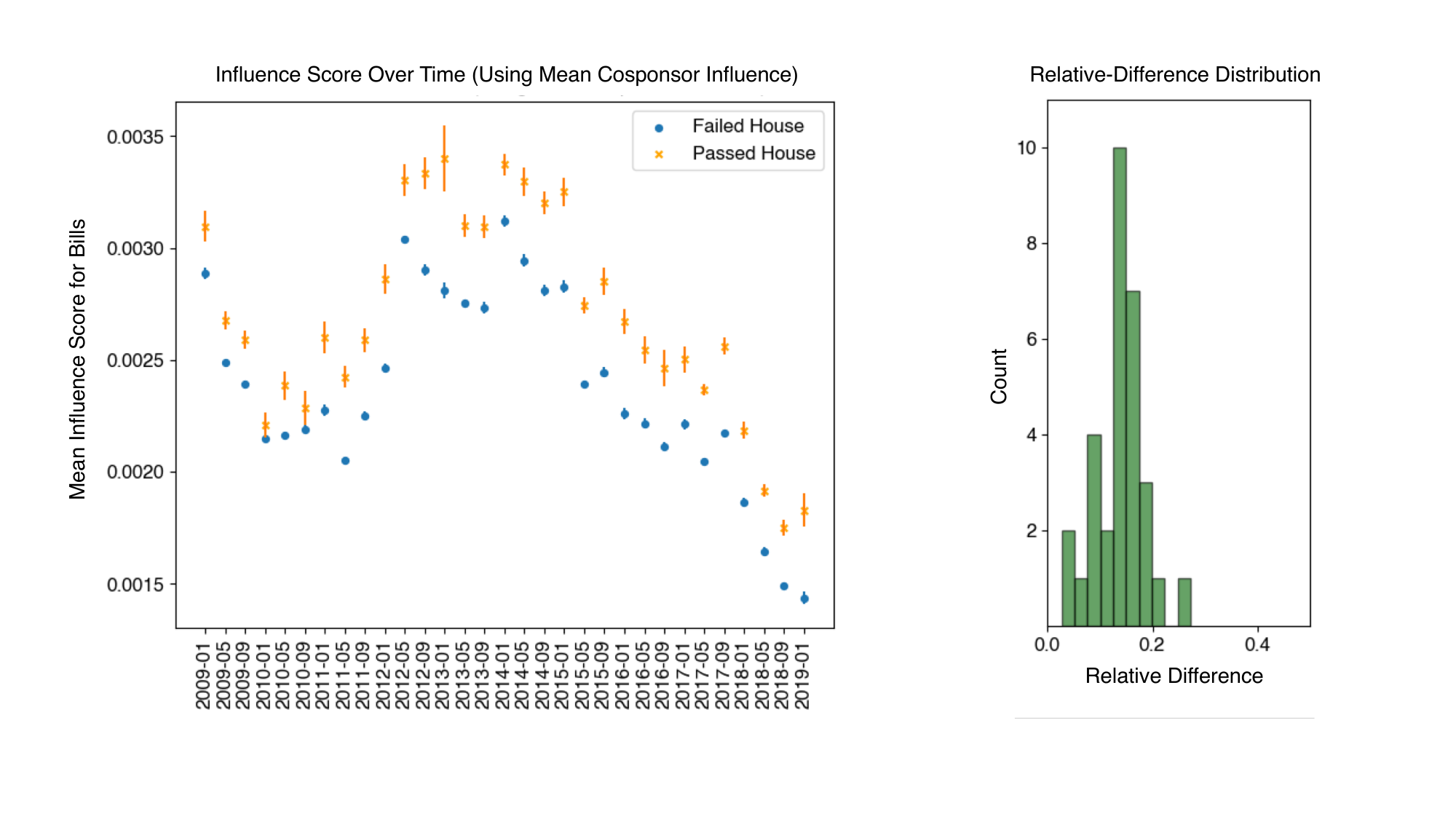}
        }
        \vspace{-1.45 cm}
        \caption{}
        \label{fig:mean6h}
    \end{subfigure}
    
    \vspace{1em} 

    \begin{subfigure}{\textwidth}
               \vspace{-.5 cm}
        \centering
        \adjustbox{max width=\textwidth, max height=0.5\textheight}{%
              \includegraphics{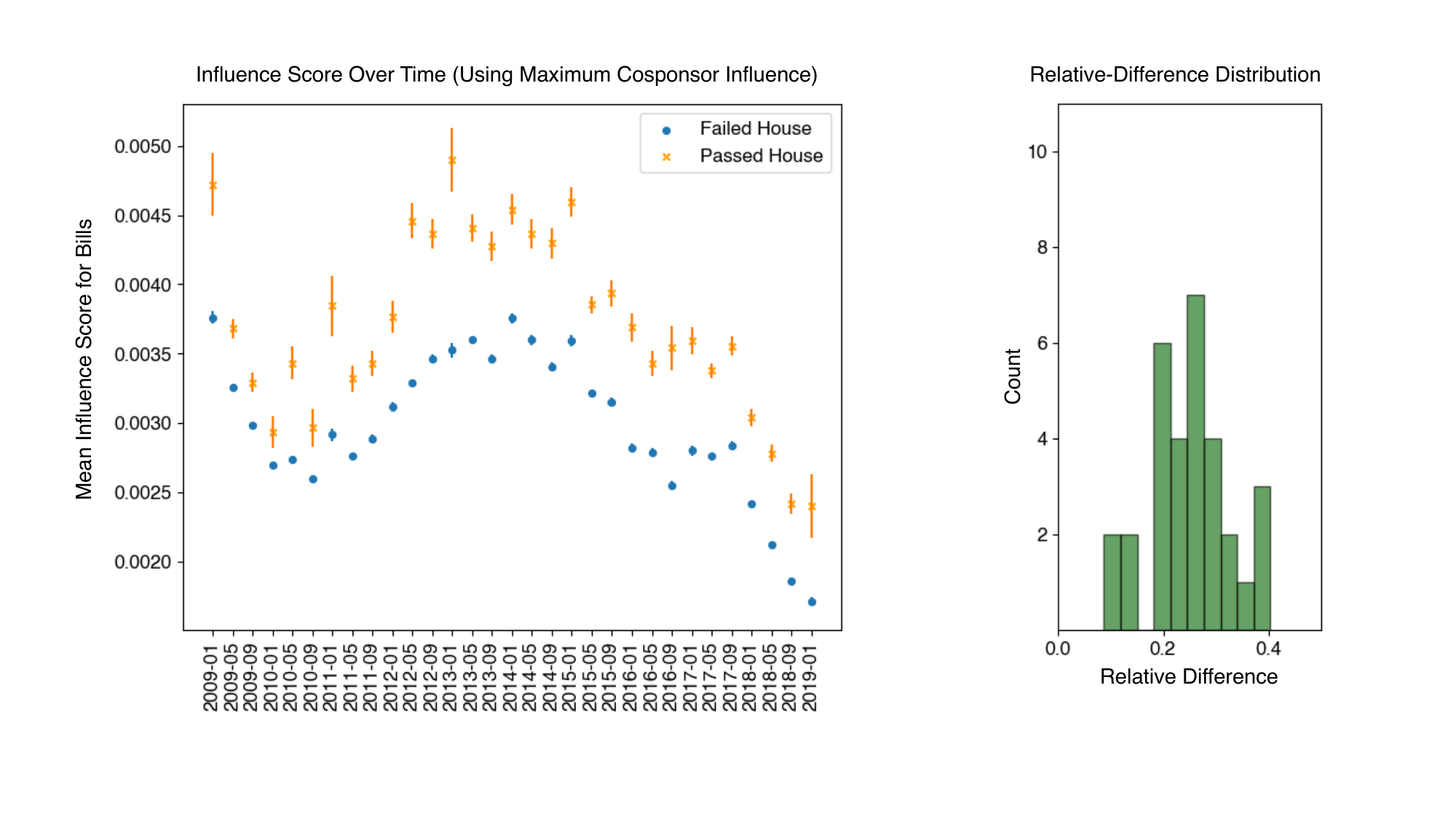}
        }
            \vspace{-1.45 cm}
        \caption{}
        \label{fig:max6h}
    \end{subfigure}
        
    \caption{
    {\footnotesize
    \textbf{Bill-influence time series and the corresponding distributions of relative differences between influence scores of bills that pass and fail in the House of Representatives.} In (a), we show bill-level influence that we calculate using the mean of a bill's cosponsors' influence scores. In (b), we show bill-level influence that we calculate using the maximum of a bill's cosponsors' influence scores. In the left panels, we show bill-level influence for 10 years, which we aggregate into nonoverlapping 4-month periods. Each point is the mean influence score of bills in the associated 4-month period starting on the date on the horizontal axis. The values of the bill-level influence scores range between $0$ and $1$.
     The orange crosses are bills that fail in the House, and the blue disks are bills that pass in the House. The error bar on each marker indicates the standard error. Our computation of influence scores incorporates an exponential decay with a half-life of 6 months. In the right panels, we plot the distributions of the relative differences between the mean influence scores of passed and failed bills. The horizontal axis gives the relative differences between the mean influence scores of bills that pass and fail in the House of Representatives. The vertical axis gives the counts of these relative differences.
  }
   } 
    \label{basicinf}
\end{figure}

In Figure \ref{basicinf}, we see that our notions of bill-level influence separate passed bills and failed bills, with passed bills having consistently larger influence scores than failed bills for all 4-month periods between 2009 and 2019 (i.e., Congresses 111--115). Therefore, our bill-level notion of influence is a relevant indicator of bill passage. In Figure \ref{basicinf}, we also see that calculating a bill-level influence score as the maximum of cosponsors' influences better differentiates between passed and failed bills than calculating bill-level influence as the mean of cosponsors' influences. In particular, observe the relative-difference distributions in the right panels. When we use a half-life of 6 months, the mean of the relative-difference distribution between bill-level influence scores of passed and failed bills is $0.248$ for maximum cosponsor influence and $0.139$ for mean cosponsor influence. We obtain qualitatively similar results for half-lives of 12 months and 24 months. For a half-life of 12 months, the relative difference between influence scores is $0.223$ for the maximum influence score and $0.122$ for the mean influence score. For a half-life of 24 months, the relative difference between influence scores is $0.203$ for the maximum and $0.109$ for the mean. See Section \ref{subsec1} for more details. Accordingly, we conclude that, on average, a bill is more likely to pass in the House if a single very influential representative cosponsors it than if several moderately influential representatives all cosponsor it. This idea relates to opinion leadership and the `influentials hypothesis', which suggests that a few people tend to be the predominant influencers of the majority of a population and the formation of public opinion~\cite{weimann1994influentials}. This phenomenon and related (but somewhat different) phenomena are relevant in various decision-making situations, such as voting and purchasing~\cite{keller2003influentials,watts2007}. We also observe a steep decline in both the mean and maximum bill-level influence scores with time, most noticeably starting in about 2015. We hypothesize that this decline is correlated with the increase in political polarization in the US during this time \cite{jacobson2018congress}. 

There are many other possible ways --- beyond using simply the mean or maximum of the cosponsor influence --- to compute a bill-level influence score from a legislator-level influence score. One possibility is to combine contributions of very influential legislators with several moderately influential legislators by employing thresholded means (e.g., by averaging the influence of all sufficiently influential legislators) or weighted means of cosponsor influence. 


\subsection{Effect of Influence Time Scale}\label{subsec1}

Our results in Section \ref{subsec5.1} suggest that using cosponsorship information to construct an influence measure of bills is informative for studying whether or not a given bill is likely to pass. In this subsection, we examine the effect of the half-life of the influence decay on how indicative our bill-level influence score (with maximum cosponsorship influence) is on bill passage. 

With this exploration, we expect to obtain insight into the time scales of legislator influence. For example, if a House member was very influential two years ago, how much does their cosponsorship of a bill now increase the likelihood that it passes? We consider half-lives of 6, 12, and 24 months. These half-lives correspond to natural time scales in Congress. Six months is half of a year, twelve months is half of a Congress, and twenty-four months is one Congress.

\begin{figure}
\centering
\adjustbox{max width=\textwidth, max height=0.5\textheight}{\includegraphics[scale=0.8]{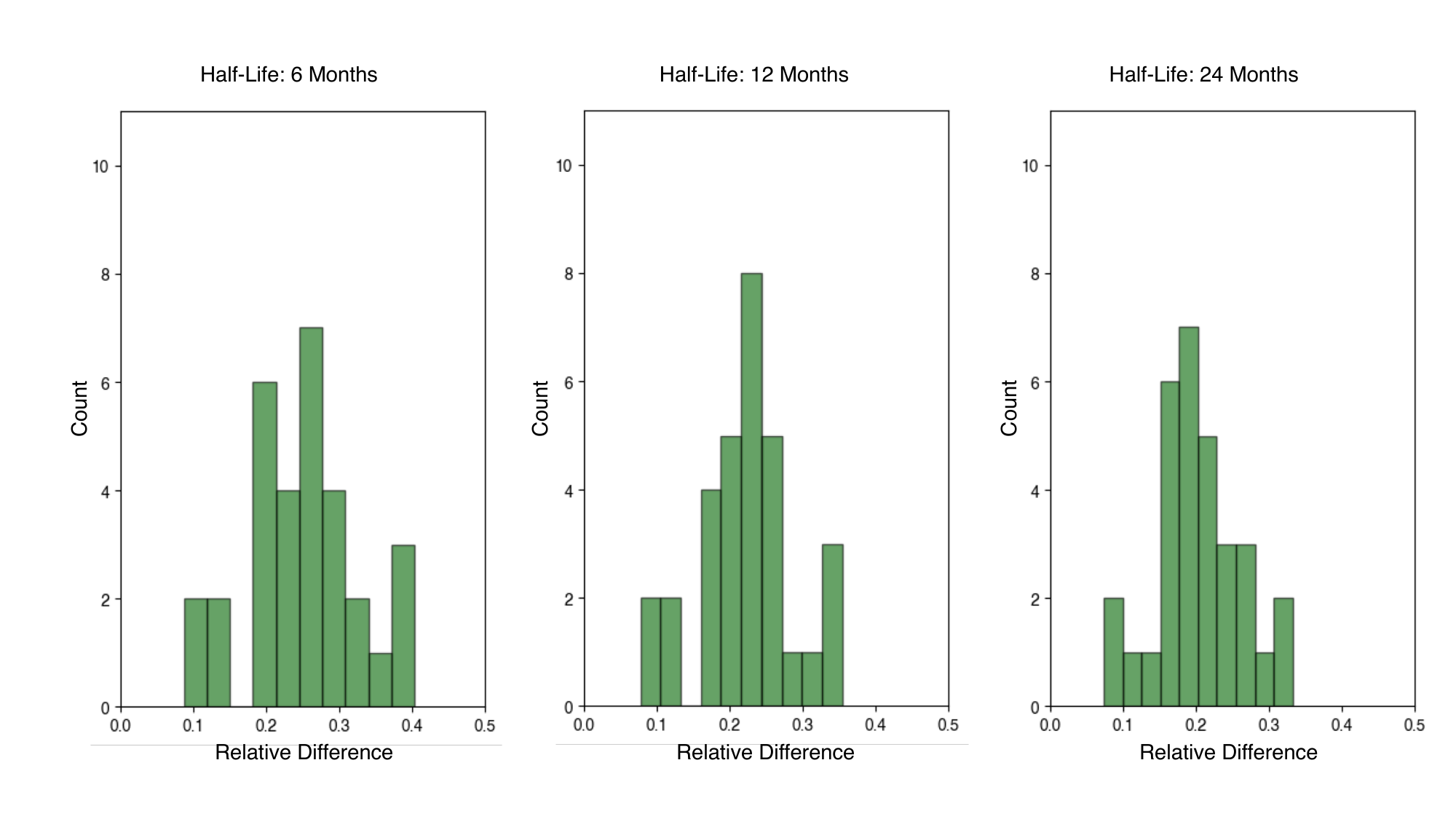}}
\caption{
{\footnotesize
\textbf{Distribution of the relative differences in influence between passed and failed bills for half-lives of 6, 12, and 24 months.} We calculate the influence using the maximum influence score of a bill's cosponsors{, and we} calculate the relative difference by normalizing the difference between the influence scores of passed and failed bills by the influence scores of the failed bills.}
}
\label{halflife}
\end{figure}

In Figure \ref{halflife}, we show the effect of the half-life of a legislator's influence decay on our ability to differentiate between passed and failed bills. For all three half-lives, we are similarly successful at distinguishing between passed and failed bills. The relative differences between the influence scores are 0.248, 0.223, and 0.203 for half-lives of 6, 12, and 24 months, respectively. The standard errors are 0.0139, 0.0119, and 0.0107, respectively. For slower decay rates, we obtain smaller differences between the influence scores of passed and failed bills. However, from a practical point of view, the differences in these influence scores are only slightly different for different half-lives. Therefore, choosing any half-life between 6 months and 24 months gives a similar ability to distinguish between passed and failed bills.


\subsection{Comparison of the Results for our Influence Measure to those for Eigenvector Centrality}\label{subsec3}

It is common to use centrality measures to examine the importances of the nodes of a network~\cite{central2025,newman2018,gomez2019centrality}. Therefore, in principle, one can calculate centralities to attempt to measure the influence of legislators. One can also interpret our legislator influence scores as centrality measures.

We now calculate the eigenvector centrality of each node (i.e., legislator). Eigenvector centrality quantifies the importance of a node based on the importances of its neighboring nodes in a network. Based on our calculations, we find that it is not a strong indicator of bill passage. In Appendix \ref{secA2}, we consider closeness centrality and strength centrality (i.e., weighted degree centrality), which are also less indicative than our influence measure of whether or not a bill passes in the House.

The eigenvector centrality of a node is given by its associated entry of the leading eigenvector of the adjacency matrix of a graph \cite{newman2018}. The eigenvector centrality of a node of a network is large when its adjacent nodes have large eigenvector centralities. Therefore, the eigenvector centrality of a node can be large even when its degree is not large. To calculate eigenvector centrality, we first divide the tensor entry $C_{\text{pass}}[t, i, j]$ \eqref{basic_matrix} (i.e., the decayed number in month $t$ of passed bills that include representatives $i$ and $j$ as cosponsors) by the tensor entry $C_{\text{tot}}[t, i, j]$ \eqref{basic_matrix} (i.e., the decayed number in month $t$ of introduced bills that include representatives $i$ and $j$ as cosponsors) to yield a tensor $D$ with entries
\begin{align}
	D[t, i, j] = \frac{C[t, i, j]_{\text{pass}}}{C[t, i, j]_{\text{tot}}} \, .
    \label{divided}
\end{align}
We then determine the largest connected component $D[t]_{\text{LCC}}$ of the network that corresponds to the month-$t$ adjacency matrix of the tensor $D$. We obtain the eigenvector centrality of representative $i$ in month $t$ by solving
\begin{align}
	D[t]{_{\text{LCC}}}x = \lambda x \, , 
\end{align}
where $\lambda$ is the leading eigenvalue of $D[t]{_{\text{LCC}}}$ and $x$ is the associated leading eigenvector. The $i$th entry of $x$ is the eigenvector centrality of representative $i$. As with our measure of legislator influence, we generate a single bill-level centrality score by calculating either the mean or the maximum eigenvector centrality of a bill's cosponsors at time $t$ (see Figure \ref{centrality}).

\begin{figure}[H]
    \centering
    
    \begin{subfigure}{\textwidth}
        \centering
        \adjustbox{max width=\textwidth, max height=0.5\textheight}{%
             \includegraphics{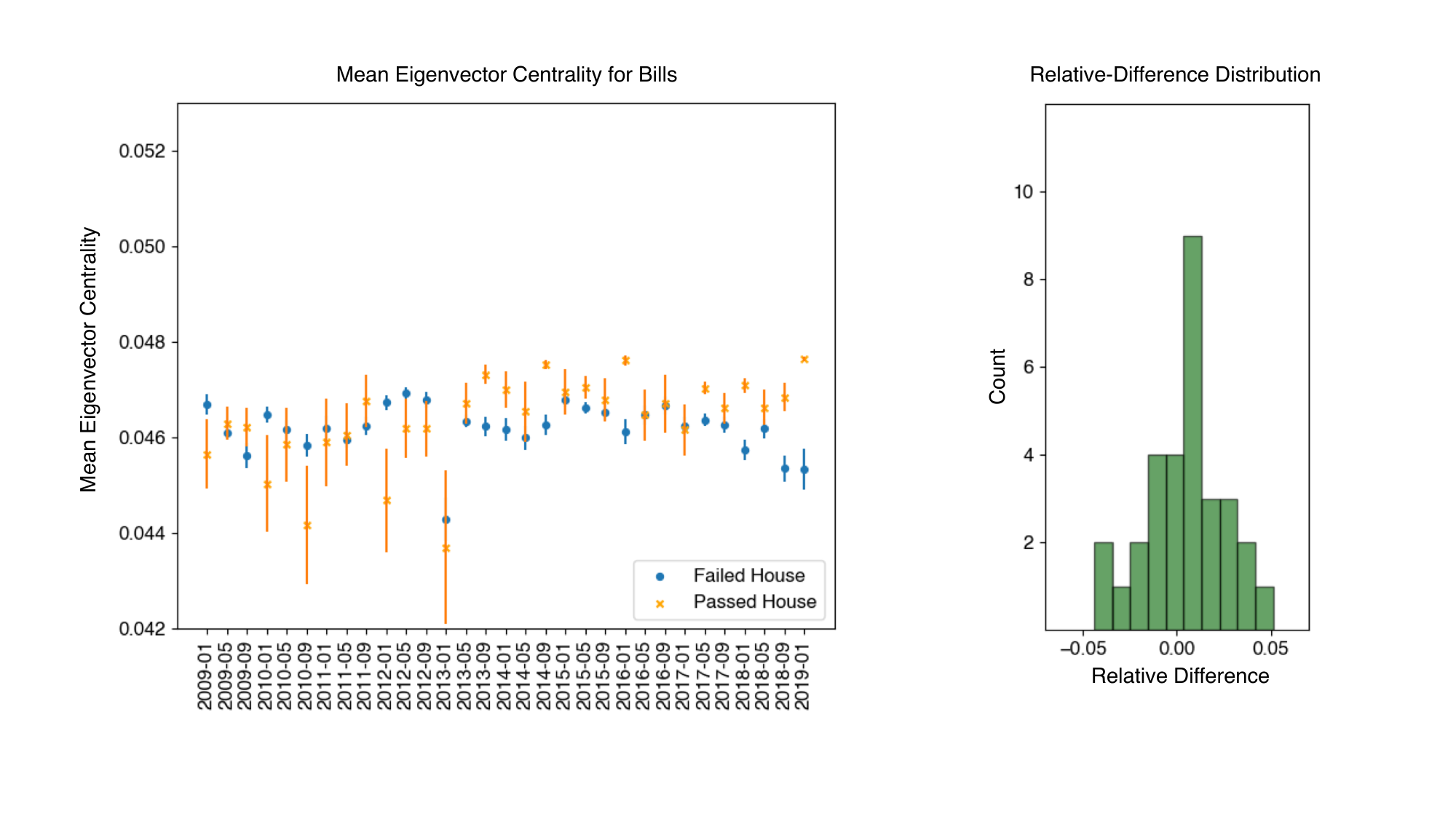}
        }
            \vspace{-1.45 cm}
        \caption{}
        \label{meancentrality}
    \end{subfigure}
    
    \vspace{1em} 

    \begin{subfigure}{\textwidth}
    \vspace{-.5 cm}
        \centering
        \adjustbox{max width=\textwidth, max height=0.5\textheight}{%
            \includegraphics{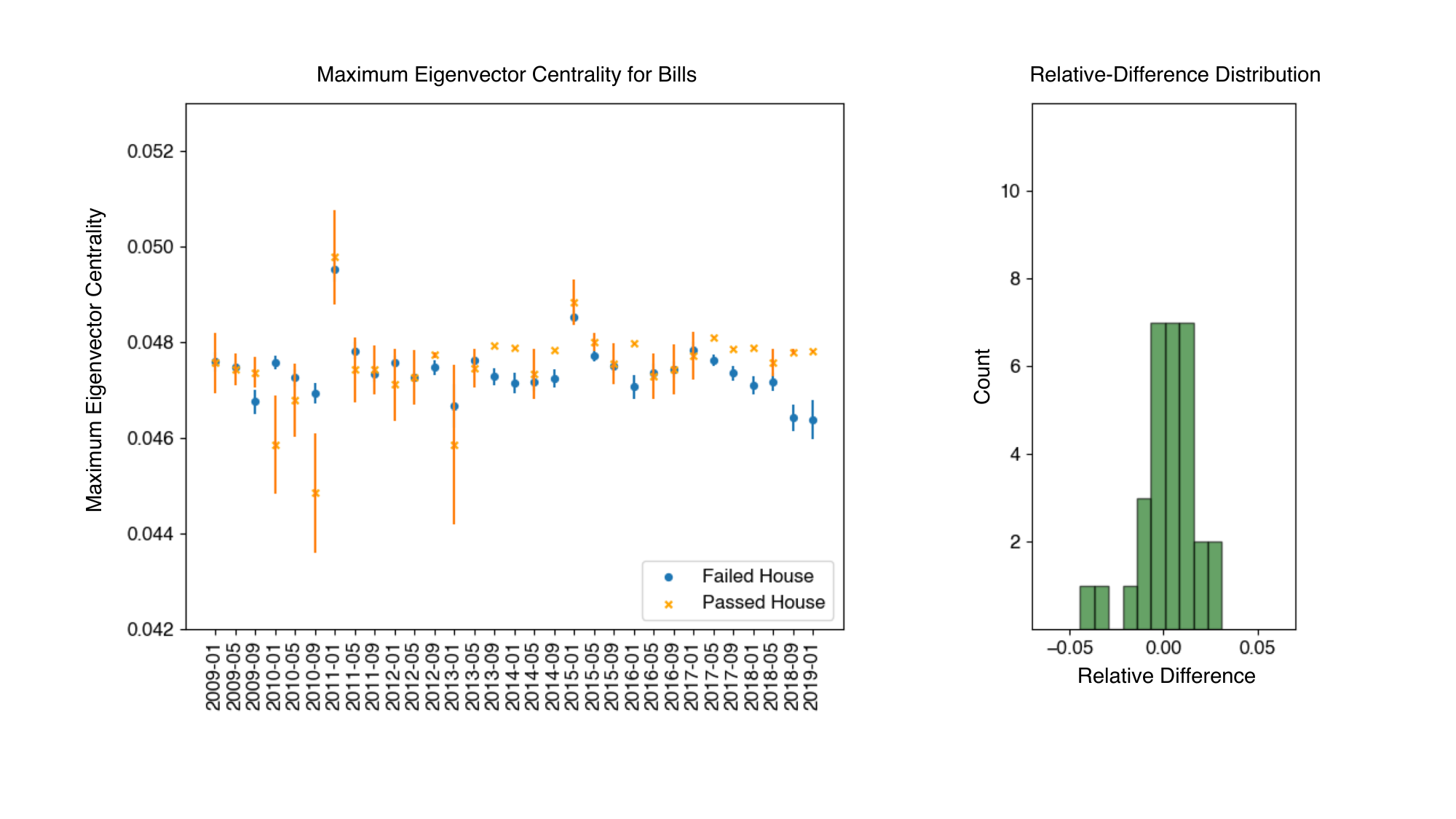}
        }
            \vspace{-1.45 cm}
        \caption{}
        \label{maxcentrality}
    \end{subfigure}
    
    \caption{
    {\footnotesize
    \textbf{Eigenvector-centrality time series and the corresponding distributions of relative differences between the eigenvector centralities of bills that pass and fail in the House of Representatives.}
    In (a), we show the bill-level eigenvector centralities that we calculate using the mean of the eigenvector centralities of a bill's cosponsors.
     In (b), we show the bill-level eigenvector centralities that we calculate using the maximum of the eigenvector centralities of a bill's cosponsors.
      In the left panels, we show bill-level eigenvector centralities for 10 years, which we aggregate into nonoverlapping 4-month periods. 
      Each point is the mean eigenvector centrality for bills in the associated 4-month period starting on the date on the horizontal axis.       
      The values of the eigenvector centralities range between $0$ and $1$.
      The orange crosses are bills that fail in the House, and the blue disks are bills that pass in the House. 
The error bar on each marker indicates the standard error. Our computation of eigenvector centralities incorporates an exponential decay with a half-life of 6 months.
 In the right panels, we plot the distributions of the relative differences between the eigenvector centralities of passed and failed bills.
 The horizontal axis gives the relative differences between the mean eigenvector centralities of bills that pass and fail in the House of Representatives. 
 The vertical axis gives the counts of these relative differences.}}
    \label{centrality}
\end{figure}

By comparing Figure \ref{centrality} to Figure \ref{basicinf} (see Section \ref{subsec5.1}), we see that eigenvector centrality is a much worse indicator of bill-passage success than our influence measure. The mean relative difference between passed and failed bills for the mean eigenvector centrality is $0.004$, and the mean relative difference for the maximum eigenvector centrality is $0.002$. These numbers are much smaller than the mean relative differences for mean influence ($0.139$) and maximum influence ($0.248$).


\section{Conclusions and Discussion}\label{sec6}

We studied the effect of bill cosponsorship relationships between legislators on bill-passage success in the US House of Representatives. To do this, we introduced a notion of influence in which legislators gain influence with each other whenever a bill that they cosponsor passes. This legislator influence, which we normalize by how often legislators cosponsor a bill, accumulates for each cosponsored bill that passes and decays exponentially with time. We leveraged our notion of legislator influence to quantify bill-level influence and study correlations between influence and the bill-passage rate.

Our calculations suggest that our influence notion, which we obtain by analyzing bill cosponsorship relationships through time, is informative about the voting outcomes of bills. For example, based on our influence measure, cosponsorship by a single very influential representative appears to be a better indicator of bill-passage success in the US House of Representatives than cosponsorship by several moderately influential representatives. We thus conclude that bill cosponsorship information yields a proxy of legislator influence, but it is only a proxy. {It is clearly desirable to conduct more granular investigations of legislator influence.}

Our work has {very} important limitations. {Our notion of legislator influence is simplistic and misses many} social nuances of influence. For instance, we treated all legislators identically, but {obviously there are many differences between individual} legislators. Different legislators are subject to different external influences and may have rather different goals. Another {significant} limitation of our analysis is that we {focused almost exclusively on} aggregated results. We purposely refained from commenting on legislator-level observations (such as which legislators are most and least influential), as we have much greater confidence in {bill-level} results than in {legislator-level} results. It {will be} valuable to build on our investigation by {thoroughly analyzing} the influence dynamics of {both} individual legislators and notable groups of legislators \cite{victor2025}. {We also appreciate that skeptical readers may feel that there is a fundamental disconnect between initially measuring a property of legislators and subsequently aggregating it to assign a property to bills. Naturally, one can also make many different choices in both the legislator-level calculations and in how (and if) one aggregates them to bill-level computations.}

There are many extensions of our study to pursue in future work. For example, it is desirable to account for factors (such as bill topics and committee assignments) beyond legislator influence that affect voting outcomes. Examining the impact of such factors will help isolate the effects of legislator influence on bill-passage success. It is also worth investigating how much our simplistic notion of influence correlates with other features of legislators, such as personal demographics, seniority {and roles within a committee (such as chairing it)}, and funding from interest groups. Additionally, it is worthwhile to examine what major political and/or social events may have influenced the temporal trends that we observed in our influence measure.

It is also important to study other links between legislator networks and their voting and cosponsorship choices. Other measures encapsulate important aspects of the relationships between legislators. One notable measure from the political-science literature is the legislative-effectiveness score (LES) \cite{volden2014legislative}. Like our influence notion, LES considers bill-voting outcomes and cosponsorship behavior. However, it differs from our approach by looking at more granular stages of the legislative process, characterizing bills by their importance, and not including a decay factor. To conduct a deeper investigation of influence dynamics in the US Congress, it is {important} to compare our legislator influence scores with LESs. It is also worthwhile to examine some of the numerous measures from network science~\cite{central2025}. In our analysis, we compared the results from our influence notion to results from three conventional centrality measures (specifically, eigenvector, closeness, and strength centralities), and we observed that those traditional centralities do not correlate strongly with bill-passage success. There are a suite of other centralities and additional network measures, and naturally one can examine other measures --- such as network efficiency~\cite{latora2001efficient}, which is concerned with how well information propagates in a network --- to see how much they correlate with bill-passage success. It also seems interesting to examine polyadic influence (e.g., see \cite{preti2024}), as perhaps multiple legislators together (e.g., legislator groups like the Squad~\cite{squad}) influence other legislators.


\backmatter


\appendix

\section{Appendix}
\label{secA2}

We now explore the influence of bill cosponsorship networks on bill-passage success using the standard centrality notions of closeness centrality (see Figure \ref{closeness}) and strength centrality (i.e., weighted degree centrality) (see Figure \ref{degree}).
 
To compute closeness centrality and strength centrality~\cite{newman2018}, we use the largest connected component of a month-$t$ Congressional cosponsorship network and its corresponding adjacency matrix $D[t]_\text{LCC}$ (see Section \ref{subsec3}). The closeness centrality of representative $i$ (i.e., node $i$) in month $t$ is
\begin{align}
	{\text{Cl}(i) = \frac{N_{{\text{LCC}} - 1}}{\sum^{N_{\text{LCC}} - 1}_{j = 1} {d_{ji}}}} \, ,  
\end{align}
where $N_{{\text{LCC}}}$ is the number of nodes in $D[t]_\text{LCC}$ and the entry $d_{ji}$ of $D[t]_\text{LCC}$ is the shortest-path distance from node $j$ to node $i$. The strength centrality of representative $i$ in month $t$ is 
\begin{align}
	s_i = \sum_{j \in {\Gamma}(i)} {W}_{ij} \, ,  
\end{align}
where $\Gamma$ is the set of neighbors (i.e., adjacent nodes) of node $i$ and $W_{ij}$ is the weight of the edge between nodes $i$ and $j$ of $D[t]_\text{LCC}$.

As with eigenvector centrality (see Figure \ref{centrality}), closeness centrality and strength centrality are worse indicators of bill-passage success than our notion of legislator influence. The mean relative difference between passed and failed bills for the mean legislator closeness centrality is $0.004$, and the mean relative difference for the maximum closeness centrality is $0.002$. The mean relative difference for the mean strength centrality is $0.004$, and the mean relative difference for the maximum strength centrality is $0.003$. We suspect that eigenvector centrality, closeness centrality, strength centrality are all unsuccessful at distinguishing between passed and failed bills for essentially the same reason, which is that the bill cosponsorship networks are very dense.

\begin{figure}[H]
    \centering
    
    \begin{subfigure}{\textwidth}
        \centering
        \adjustbox{max width=\textwidth, max height=0.5\textheight}{%
            \includegraphics{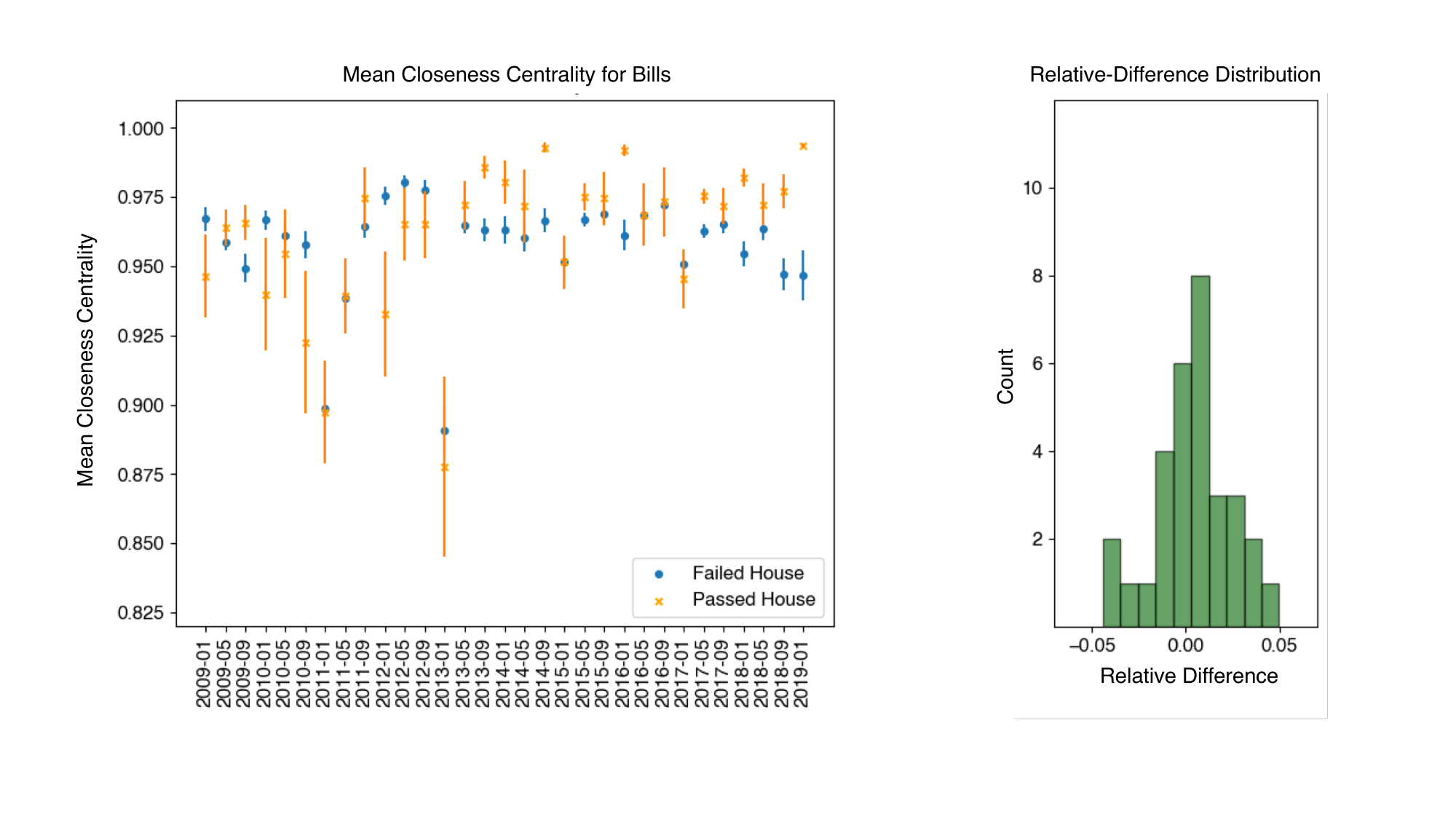}
        }
            \vspace{-1.45 cm}
        \caption{}
        \label{meancloseness}
    \end{subfigure}
    
    \vspace{1em} 

    \begin{subfigure}{\textwidth}
       \vspace{-.5 cm}
        \centering
        \adjustbox{max width=\textwidth, max height=0.5\textheight}{%
            \includegraphics{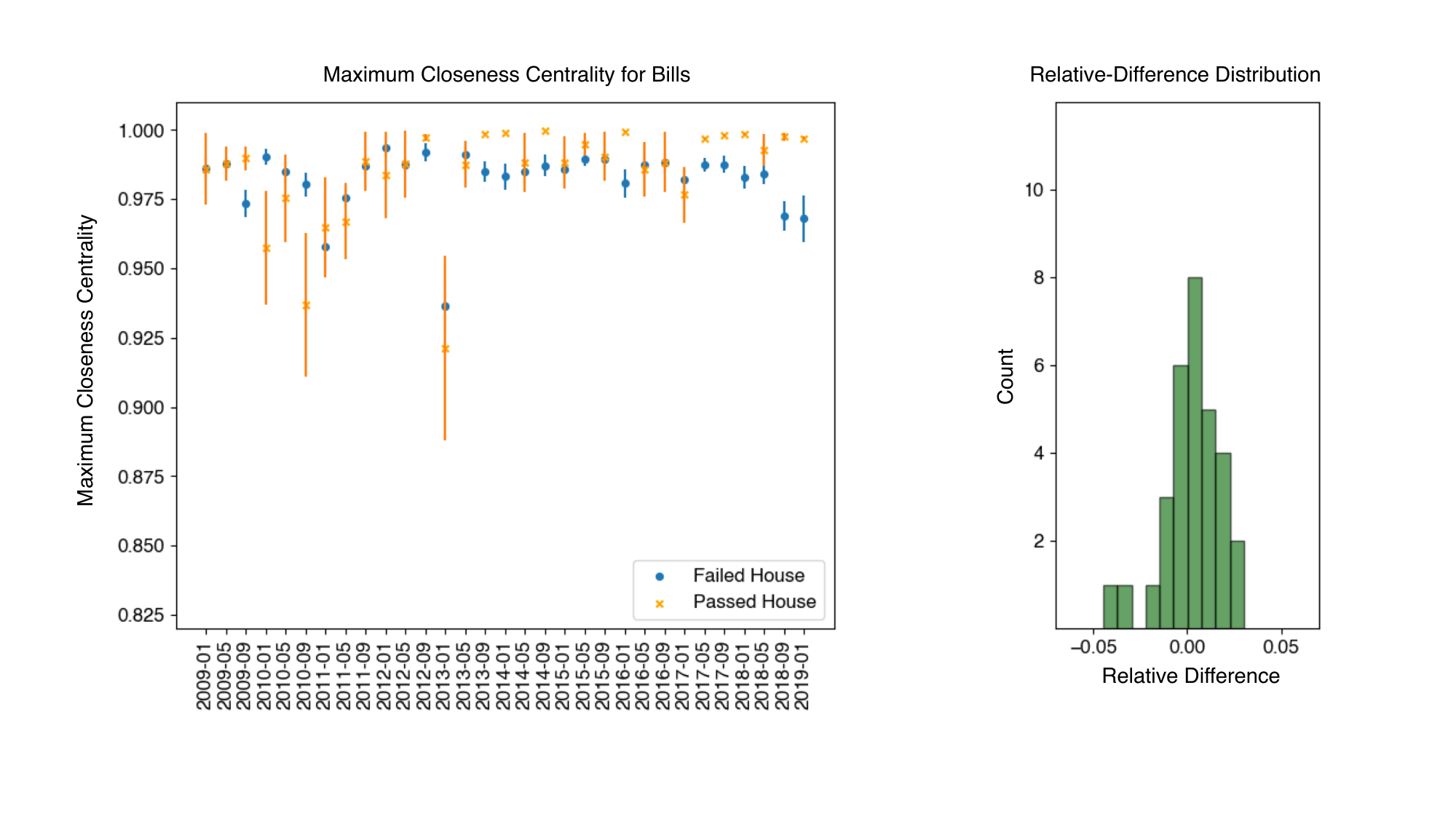}
        }
            \vspace{-1.45 cm}
        \caption{}
        \label{maxcloseness}
    \end{subfigure}
    
    \caption{
    {\footnotesize
    \textbf{Closeness-centrality time series and the corresponding distributions of relative differences between the closeness centralities of bills that pass and fail in the House of Representatives.}
        In (a), we show the bill-level closeness centralities that we calculate using the mean of the closeness centralities of a bill's cosponsors.
     In (b), we show the bill-level closeness centralities that we calculate using the maximum of the closeness centralities of a bill's cosponsors.
      In the left panels, we show bill-level closeness centralities for 10 years, which we aggregate into nonoverlapping 4-month periods. Each point is the mean closeness centrality of bills in the associated 4-month period starting on the date on the horizontal axis. 
           The values of the closeness centralities range between $0$ and $1$.
      The orange crosses are bills that fail in the House, and the blue disks are bills that pass in the House. 
The error bar on each marker indicates the standard error. Our computation of closeness centralities incorporates an exponential decay with a half-life of 6 months.
 In the right panels, we plot the distributions of the relative differences between the closeness centralities of passed and failed bills.
 The horizontal axis gives the relative differences between the mean closeness centralities of bills that pass and fail in the House of Representatives.
 The vertical axis gives the counts of these relative differences.}}
    \label{closeness}
\end{figure}

\begin{figure}[H]
    \centering
    
    \begin{subfigure}{\textwidth}
        \centering
        \adjustbox{max width=\textwidth, max height=0.5\textheight}{%
           \includegraphics{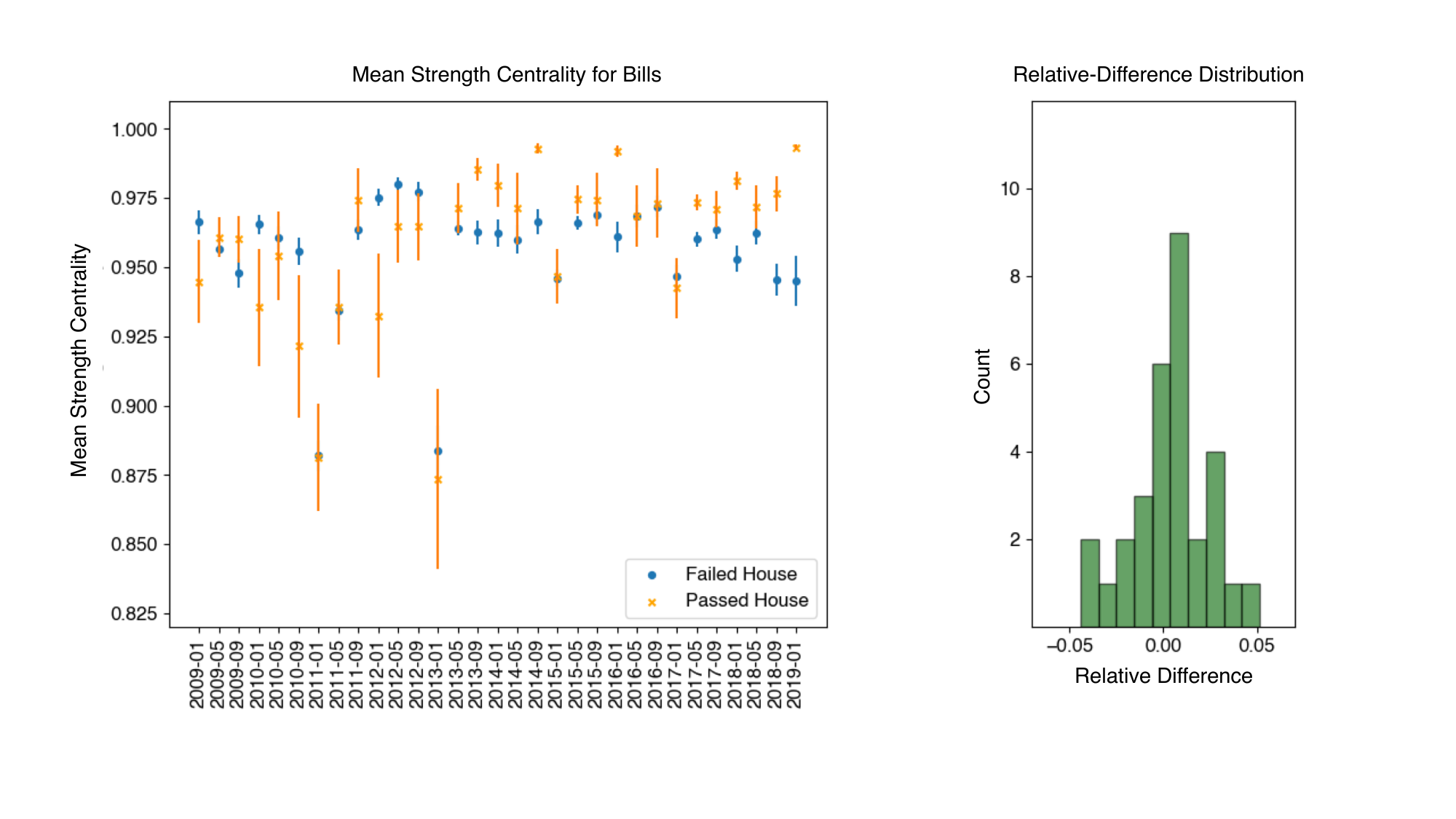}
        }
            \vspace{-1.45 cm}
        \caption{}
        \label{meanstrength}
    \end{subfigure}
    
    \vspace{1em} 

    \begin{subfigure}{\textwidth}
        \vspace{-.5 cm}
        \centering
        \adjustbox{max width=\textwidth, max height=0.5\textheight}{%
           \includegraphics{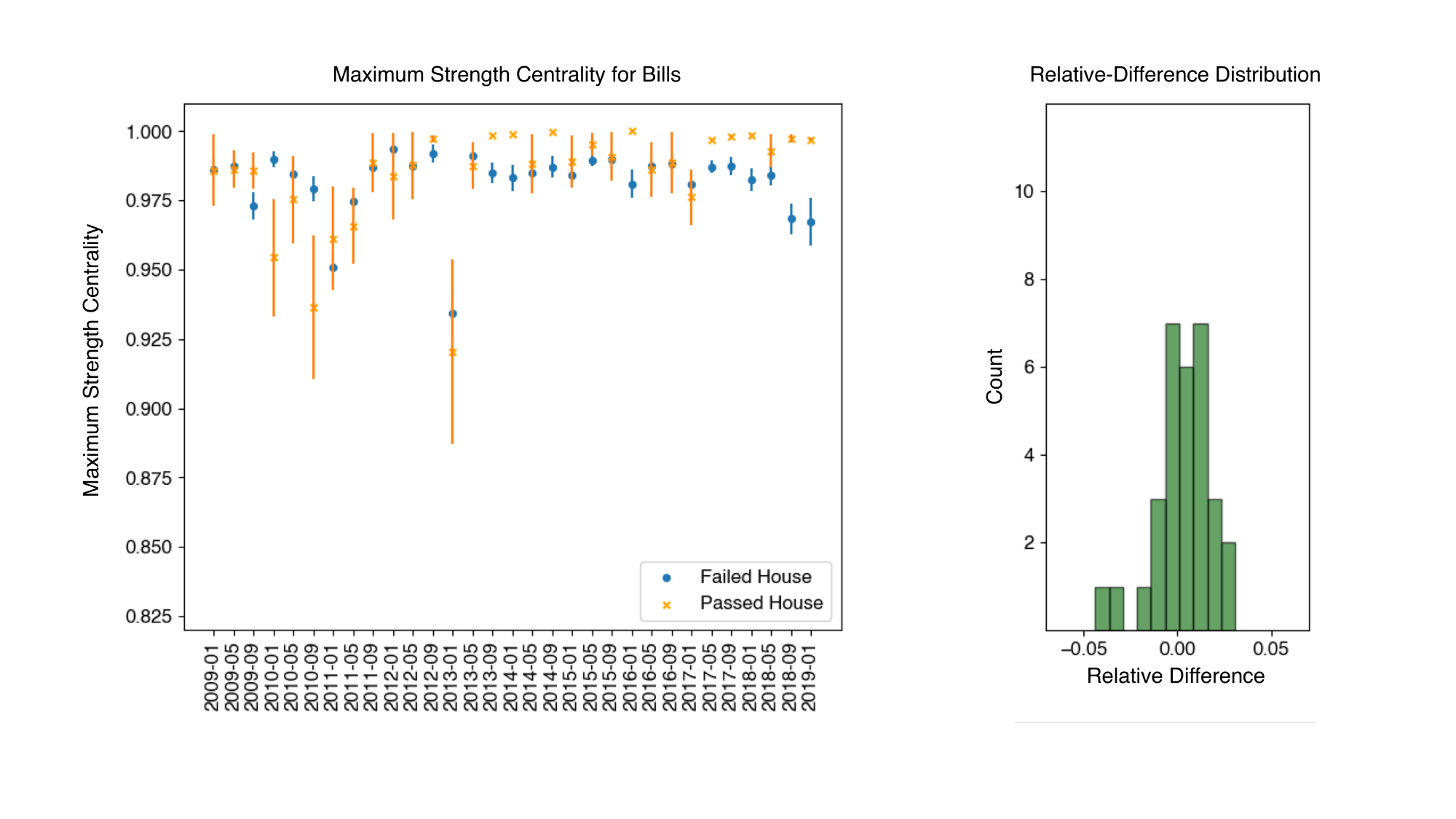}
        }
            \vspace{-1.45 cm}
        \caption{}
        \label{maxstrength}
    \end{subfigure}  
    \caption{
    {\footnotesize
    \textbf{Strength-centrality time series and the corresponding distributions of relative differences between the strength centralities of bills that pass and fail in the House of Representatives.}
        In (a), we show the bill-level strength centralities that we calculate using the mean of the strength centralities of a bill's cosponsors.
     In (b), we show the bill-level strength centralities that we calculate using the maximum of the strength centralities of a bill's cosponsors.
      In the left panels, we show bill-level strength centralities for 10 years, which we aggregate into nonoverlapping 4-month periods. Each point is the mean strength centrality for bills in the associated 4-month period starting on the date on the horizontal axis.  
           The values of the strength centralities range between $0$ and $1$.
      The orange crosses are bills that fail in the House, and the blue disks are bills that pass in the House. 
The error bar on each marker indicates the standard error. Our computation of strength centralities incorporates an exponential decay with a half-life of 6 months.
 In the right panels, we plot the distributions of the relative differences between the strength centralities of passed and failed bills.
 The horizontal axis gives the relative differences between the mean strength centralities of bills that pass and fail in the House of Representatives. 
  The vertical axis gives the counts of these relative differences.}}
    \label{degree}
\end{figure}


\section*{Declarations} 


\subsection{Availability of data and material}

Our code is publicly available at \url{https://github.com/sarahsotoudeh/LegislativeInfluence}. 


\subsection{Competing interests}

We have no competing interests.


\subsection{Funding}

SK acknowledges financial support from the UC Presidential Postdoctoral Fellowship.


\subsection{Authors' contributions}

SS and SK conceived and conceptualized the study. SS performed the analysis and wrote the initial draft of the paper. SS, MAP, and SK reviewed and extensively edited the manuscript, determined what additional analysis was necessary, and produced the final version of the manuscript. All authors read and approved the final manuscript.


\subsection{Acknowledgements}

We thank Martin Gilens, Zachary C. Steinert-Threlkeld, Jennifer Nicoll Victor, and {three} anonymous referees for helpful comments. 






\end{document}